\documentclass{aastex62}

\newcommand\aastex{AAS\TeX}

\received{}
\revised{}
\accepted{}
\submitjournal{ApJ}

\shorttitle{\aastex\ Radial acceleration in galaxies}
\shortauthors{A. Maeder}

\begin{document}

\title{The acceleration relation  in galaxies and scale invariant dynamics:  another challenge for dark matter}


\correspondingauthor{Andre Maeder}
\email{andre.maeder@unige.ch}

\author[0000-0002-0786-7307]{Andre Maeder}
\affiliation{Geneva Observatory \\
chemin des Maillettes, 51 \\
CH-1290 Sauverny, Switzerland}

\begin{abstract}
We examine the radial acceleration relation (RAR)
between the centripetal acceleration $g_{\mathrm{obs}}$  and  
the gravity due to the baryons  $g_{\mathrm{bar}}$ in galaxies  \citep{McGaugh16,Lelli17}. 
 Below about $10^{-10}$ m s$^{-2}$, 
the  RAR deviates from the 1:1 line,
$g_{\mathrm{obs}}$   being much larger than $g_{\mathrm{bar}}$.
 The RAR is followed by late and early  type galaxies,  and also by the
dwarf spheroidals where the deviations from the 1:1 line are the largest ones.  These deviations  are  currently attributed  to dark matter. 

We show that the   scale invariant theory, with the assumption of the scale invariance of   the empty space,
 correctly predicts  the observed  deviations in the acceleration relation.
 The large deviations (up to a factor 400) and the  flattening of the acceleration relation 
 observed for the dwarf spheroidal galaxies  are also well described.
 The presence of dark matter  is no longer  necessary in the scale invariant context, which also accounts  why 
 dark matter  usually appears to dominate  in  regions  of late type galaxies  with low baryonic  gravities.

\end{abstract}


\section{Introduction: the context} \label{sec:intro}

The relation between  the dynamics of galaxies and the    distributions of their  visible matter   
opened  a large variety of studies 
on the dark matter and gravitation \citep{SRubin01}. The radial acceleration relation (RAR)
 compares the radial acceleration traced by the rotation curves of galaxies 
and the acceleration expected from the observed distribution of baryons \citep{McGaugh04}.  It is also often expressed in terms of a 
 mass discrepancy acceleration relation (MDAR).  Recent extensions of this relation to different 
types of galaxies have been performed by \citet{McGaugh16} and 
\citet{Lelli17}. Their  conclusion  is that the dark matter (DM) distribution is fully determined by that of the baryons or vice-versa.
A number of authors have interpreted this relation in the context of the $\Lambda$CDM models 
of galaxy formation, in terms of different  mass-dependent density profiles of 
 DM haloes \citep{Dicintio16},  of  simulations of galaxy formation matching the velocities and the scaling relations  \citep{Santos16}, 
 in particular if  stellar masses and sizes are closely
 related to the masses and sizes of their DM haloes \citep{Navarro17}.  \citet{Keller17} show that the account for the hot outflows from supernovae  at high $z$ improves the comparisons
  by producing a baryon depletion, while the effects of the AGN feedback is considered by \citet{Ludlow17}.
  The role of the galaxy-halo connexion is also emphasized by \citet{Desmond17}.   Attempts to explain  the observed relation 
in the context of modified gravity  (MOND) have been made by
\citet{Milgrom16} and \citet{Li18}.

New cosmological models have recently been proposed, 
which include the specific hypothesis that the macroscopic empty space is invariant to 
 a scale transformation, a property absent from general relativity (GR) with a cosmological constant,
  but present in the Maxwell equations \citep{Maeder17a}.
These models are based on the general field equation in Weyl's geometry obtained by \citet{Dirac73} and \citet{Canu77}, which includes also 
the possibility of scale invariance in addition to the general covariance of GR.
 These models predict an accelerated cosmic expansion without advocating the existence of some unknown form of 
  dark energy. Several comparisons between models and observations have been performed: on the distance vs. redshifts, 
  the $m-z$ diagram, the $\Omega_{\Lambda}$ vs. $\Omega_{\mathrm{m}}$ plot, the age vs. $H_0$, the expansion rate $H(z)$ vs.
  redshifts $z$ and on the redshift of the transition 
  from braking to the present acceleration of the cosmic expansion. 
  A critical analysis of the CMB temperatures 
  as function of  $z$ from CO molecules in DLA is also supporting these cosmological models \citep{Maeder17b}.

 The   scale invariant  models also have consequences for  weak gravitational fields. A geodesic 
 equation in the scale invariant  framework was  obtained by \citet{BouvierM78}.
  Applied to  a weak field,  the relation corresponding to  Newton's law  was derived. 
  It contains an additional  small acceleration term proportional to the velocity,
 particularly significant in low gravity and low density regions \citep{MBouvier79,Maeder17c}.
 This leads to an energy relation, which for clusters
 of galaxies provides  mass determinations much  smaller than usual, letting little or  no room for dark matter.
 A detailed study of the two-body problem has also been made in the scale invariant framework. This allows, alike the MOND theory 
 \citep{Milgrom83}, but in a more general context, to account for the flat 
   rotation curve of the Milky Way  and galaxies \citep{SRubin01,Sohn17}, and also why no dark matter is
  suggested  in forming galaxies at redshifts around z =2 \citep {Wuyts16,Genzel17,Lang17}.

 Here, we want to check whether the RAR or MDAR
can  be interpreted  within the scale invariant  framework.
 In Section 2, after  
  an additional demonstration of the gauging condition,
  the acceleration relation is studied in the scale invariant context.
  In Section 3, the observations and  predictions of the scale invariant theory are compared. Section 4 presents a 
  discussion about dark matter 
  and  on some possibilities to distinguish between the $\Lambda$CDM and the scale invariant theories.
   Conclusions are given in Section 5.

  \section{The radial acceleration relation in the scale invariant theory}  \label{theory}
  

  \subsection{The models and the gauging conditions}  \label{gauge}

The models studied  here are based on the theoretical developments made by 
\citet{Weyl23,Eddi23,Dirac73,Canu77} and \citet{BouvierM78} for a gravitation theory including scale invariance in addition to the
general covariance. A scale transformation of the line element 
  $ ds'$ of GR is expressed by
  \begin{equation}
  ds'= \lambda(x^{\mu}) ds \,
  \label{lambda}
  \end{equation}
\noindent  
   where $ ds $ is a new line element. Weyl’s Geometry, instead of Riemann Geometry, offers the
  consistent framework, in which scale invariance can be expressed in addition to the general covariance of GR.
  Dirac and  later  Canuto et al. have  expressed in the scale invariant framework an action principle, which 
   includes a matter Lagrangian as a coscalar of power -4. In the integrable  Weyl’s space,
    Dirac and Canuto et al. have shown that the scale factor remains undetermined without any other constraints. 
    Later, they were fixing the scale factor by some external considerations based on the large number hypothesis 
    (instead of that we fix it by an assumption on the vacuum property).  From the differential of  the action, 
    a  generalization of the Einstein field equation satisfying both general covariance and scale invariance was obtained.
    This is  Equation (7) used in the paper by \citet{Maeder17a}. 
    This field equation can also be derived in the step by step developments of the so-called cotensor calculus, 
    which  is the basic mathematical tool of Weyl’s geometry, like tensor calculus  in Riemann Geometry.
    Scale covariant derivatives, modified Christoffel symbols, the Riemann-Christoffel tensor also have their
  scale covariant counterparts.  A short summary on cotensors has been given by \citet{Canu77}.
     
The contribution by Bouvier and Maeder was to obtain the geodesic equation,  
 expressing the motion of a free particle from a minimum action expressing the shortest distance between two points 
 in the integrable Weyl’s space.  
 A derivation of the geodesic equation from the Equivalence Principle was further given by \citet{MBouvier79}. They
 also showed that this geodesic equation leads to a slightly modified Newton’s equation, see below Equation (\ref{Nvec})  \citep{Maeder17c}.
  Thus, both the general field equation and the geodesic equation in the scale invariant context 
  used in the previous papers have been thoroughly studied and analyzed in the cotensor framwork and from an action principle.

  In Equation (\ref{lambda}), $\lambda$ is the scale factor, which for reasons of homogeneity and isotropy  only depends on time.
  The ''new'' general field and geodesic  equations 
  contain   terms depending on the coefficient of metrical connection $\kappa_{\nu}$,
  \begin{equation}
\kappa_{\mu} \, = \, -\frac{\partial}{\partial x^{\mu}  } \, \ln \lambda \, .
\label{kappa}
\end{equation}
\noindent
and its derivatives.   The geodesic equation writes,
\begin{equation}
\frac{du^{\rho}}{ds}+ \Gamma^{\rho}_{\mu \nu} u^{\mu} u^{\nu} -\kappa_{\mu}u^{\mu} u^{\rho}+ \kappa^{\rho} = 0 \, ,
\label{geod}
\end{equation}
\noindent
with the velocity  $u^{\mu} \, = \, dx^{\mu}/ds$.  It contains two additional terms depending on $\kappa_{\mu}$.
Interestingly enough, for $\kappa_{\mu}$ given by
Equation (\ref{kappa}), any vector undergoing a parallel displacement along a closed curve has its length unchanged in Weyl's space,
alike in Riemann space, while this would not be the case for other forms of  $\kappa_{\mu}$. 
Other properties of Weyl's geometry  define a consistent framework for gravitation \citep{BouvierM78}.

 It is often not realized  that Einstein's  field equation  in GR is scale invariant, when the cosmological constant $\Lambda_{\mathrm{E}}$
 is equal to zero. 
 However, the point is that the property of scale invariance is no longer  verified, 
  when the cosmological constant is different from zero \citep{Bondi90}.
 The specific hypothesis we have made to the fix the gauge $\lambda$ \citep{Maeder17a} is 
 that the macroscopic empty space should be invariant to 
 a scale transformation even if the cosmological constant is different from zero.
 In this respect, we recall that the Maxwell equations are also scale invariant in the empty space.
   When this 
  hypothesis is expressed  in the general scale invariant field equation,
 it  leads to  two differential equations  between the scale factor $\lambda(t)$, its derivatives 
  and the cosmological constant $\Lambda_{\mathrm{E}}$ \citep{MBouvier79,Maeder17a},
  \begin{equation}
\  3 \, \frac{ \dot{\lambda}^2}{\lambda^2} \, =\, \lambda^2 \,\Lambda_{\mathrm{E}}  \, , 
\label{diff1}
 \end{equation}
 \begin{equation}
\frac{\ddot{\lambda}}{\lambda} \, = \,  2 \, \frac{ \dot{\lambda}^2}{\lambda^2} \, ,
\label{diff2}
\end{equation}
\noindent  
or some combination of the two. 
The solution of these equations  gives  $\lambda$ varying like $ 1/t$.
Thanks to  these relations  $\Lambda_{\mathrm{E}}$ can be eliminated
from the cosmological equations and  a fully determined system of equations is obtained for the chosen metric. 
 Interestingly enough, the above  relations  bring major simplifications  in the  cosmological equations by \citet{Canu77}.\\

 Now, let us examine more  closely  the meaning of the above  relations 
 between the cosmological constant and the scale factor $\lambda$.
  The cosmological constant  
  is related by some constant factor to the energy density $\varrho_v$ of the empty space, see for example \citet{Carr92},
  \begin{equation}
  \Lambda_{\mathrm{E}} \, = \, 8 \pi \, G \, \varrho_v \, ,
  \label{vac}
  \end{equation}
  \noindent
 with $c=1$.
  Let  $\ell'$  be a  constant line element in the space of GR. 
In the scale invariant space, the line element becomes
 $ \ell \, = \frac{\ell'}{\lambda(t)}$,  
  where $\lambda$ is only a function of the cosmic time $t$, as said above.  
  The possible variations of the scale factor $\lambda$ may  contribute to
 the  energy density present in the empty space. Thus, 
 if  $\lambda(t)$ varies, the  energy density present in the empty space will depend on
$
\dot{\ell}^2 \, = \, \ell'^2 \, \frac{\dot{\lambda}^2}{\lambda^4}.$
 If there is no other source of energy in the macroscopic empty space, 
  the cosmological constant  
is  thus  proportional to $\dot{\ell}^2$,
\begin{equation}
\Lambda_{\mathrm{E}} \, = \, const. \frac{\dot{\lambda}^2}{\lambda^4} \, ,
\label{LE}
\end{equation}
\noindent
which compares with Equation (\ref{diff1}).
The cosmological constant $\Lambda_{\mathrm{E}} $
is a true constant and this implies,
\begin{equation}
\frac{d \Lambda_{\mathrm{E}}}{dt} \, =  \, 0 \, \quad \Longrightarrow \quad
\frac{d \dot{\ell}^2}{dt^2}=
2 \dot{\ell} \, \ddot{\ell} \, = \, \ell'^2 \left( \frac {\dot{\lambda} \ddot{\lambda}}{\lambda^4} -
2 \, \frac{\dot{\lambda}^3}{\lambda^5}\right) \, = \, 0 \,  .
\label{d1}
\end{equation}
\noindent
Expressing $\ddot{\ell}$, we get 
$
\ddot{\ell} \, =  \, \frac{\ell'}{\lambda} \left(\frac{\ddot{\lambda}}{\lambda} - 
2 \, \frac{\dot{\lambda}^2}{\lambda^2}\right)$.
The second derivative of $\ell$ should be zero in order to satisfy Equation (\ref{d1}) for  $\dot{\ell} \neq \, 0$, thus 
\begin{equation}
 \ddot{\ell} \, =  \,0, \quad \Longrightarrow \quad 
\frac{\ddot{\lambda}}{\lambda} \, = \,  2 \, \frac{ \dot{\lambda}^2}{\lambda^2} \, ,
\label{d2}
\end{equation}
\noindent
in agreement with Equation (\ref{diff2}). We see that the gauging conditions, independently obtained 1)
from the  hypothesis of  scale invariance of the   empty space  and   2)  from the constancy of
the energy density of the  empty space are both giving  identical results.
This confirms the consistency of the adopted gauging condition. In GR, the cosmological constant, and thus the density
of the empty space,  is  independent of the material  content of the Universe.  This also applies
to the conditions (\ref{diff1}) and (\ref{diff2}) 
expressing  the energy--density of the empty space in terms of the scale factor. 
 In this framework,
  the only significant  energy  in the large scale empty space
is that  resulting from the gauge variations. 
Indeed, we consider this applies to large scales,
even if this  is not necessarily true at   the  quantum level,
 in the same way as we may use   Einstein theory at large scales, even if we cannot do it at the quantum level.



With the Robertson-Walker metric,
the above gauging conditions (\ref{diff1}) and (\ref{diff2})  enabled us to express the cosmological equations
(29) - (31) in \citet{Maeder17a}. 
   The resulting  models predict an acceleration of the cosmic
 expansion and satisfy various cosmological  tests.
 

\subsection{Some basic useful relations in the scale invariant context}  \label{recall}

 The above  scale invariant equation of the geodesics  applied
  in the weak field approximation  is leading to a modified form of the Newton equation \citep{MBouvier79,Maeder17c},
 \begin{equation}
 \frac {d^2 \bold{r}}{dt^2} \, = \, - \frac{G \, M}{r^2} \, \frac{\bold{r}}{r}   + \, \kappa(t)\,  \frac{d\bold{r}}{dt} \, ,
\label{Nvec}
\end{equation}
\noindent
where $\kappa(t) \, = \, 1/t $  with $t$ being the cosmic time. In addition to the 
 usual gravitational attraction it contains a small    acceleration term proportional the velocity, particularly significant in low gravity and 
 low density media (Section \ref{gdens}). The additional term results from the cosmological constant, which  is related to the scale factor 
 $\lambda$ and its first derivative by Equation (\ref{diff1}).
For a constant scale factor $\lambda$, there is no  acceleration term.
The law of conservation of angular momentum $L$
 is different from the usual one, 
\begin{equation}
\kappa(t) \,  r^2 \, \dot {\vartheta} \, = \, L  \, = \, \mathrm{const.}
\label{ang}
\end{equation}
\noindent
It is a scale invariant quantity. 
The equation of motion  (\ref{Nvec}) has been applied to the two-body problem 
and  an equivalent form to Binet equation has been found. The 
orbits  $r(\vartheta)$  belong to the usual family of conics, with in addition a slight outward expansion motion,
 \begin{equation}
 r  \, = \, \frac{r_0}{1 + e \cos(\vartheta)}, \quad
 \mathrm{with} \quad r_0 \, = \, \frac{L^2}{G \, M \, \kappa^2(t)} \, ,
 \label{sol}
 \end{equation}
\noindent
where  $r_0$ is the radius of the circular orbit. The eccentricity $e$ is scale invariant, while $r_0$ varies like $t$.
 The circular velocity in the case of the two-body problem is according to Equation (\ref{ang}) 
$ v_{\mathrm{c}} \, = \, r_0 \, \dot{\vartheta} \, = \, \frac{L}{\kappa(t) \, r_0}$,
which is also a scale invariant quantity. Now, with the expression (\ref{sol}) for $r_0$, one obtains
\begin{equation}
v^2_{\mathrm{c}} \,  = \frac{G \, M}{r_0} \, .
\label{velocity2}
\end{equation}
\noindent
The expression relating the circular velocity and the gravitational potential
is  valid. Consistently, the gravitational potential is also 
a scale invariant quantity, since we have both $r'= \lambda r$ and $M'= \lambda M$, as already seen in \citet{Maeder17a}.
For an elliptical motion, the semi-major axis $a$  and the period $P$ also behave like $\dot{a}/a =\dot{P}/P = 1/t$.
 These relations  illustrate the main properties of the motions.
 The additional outward acceleration term in Equation (\ref{Nvec}) does not produce
 a change  of the circular velocity, but  just an extension of the orbit and  a corresponding reduction of the angular velocity. 

\subsection{Energy relation in the scale invariant context}  \label{enrel}

A gravitational system governed by Equation (\ref{Nvec}) does not reach  
a state of perfect  equilibrium  in the long term, since the additional term produces secular variations of the orbits.
The additional term  in this equation does not derive from a potential,
thus the conditions are different from those of the usual virial theorem. Since the variations of the orbits due to the additional
term are currently slow, a gravitational system may find a state of quasi-equilibrium. Thus,
we may find an expression which relates the different forms of energy present.
Let us consider the ensemble of $N$ stars moving in a galaxy.
According to Eq.(\ref{Nvec}), the acceleration of a star $i$  interacting with another one of mass $m_j$ is
\begin{equation}
 \frac{dv_{ij}}{dt} \, = \,-\frac{G \, m_j}{r^2_{ij}} + \kappa(t) \, v_{ij}\, \, ,
\end{equation}
\noindent
where $r_{ij}$ is the distance between objects $i$ and $j$ and $v_{ij} = \frac{dr_{ij}}{dt}$. 
First, we multiply  this equation  by $v_{ij}$ and get 
\begin{equation}
  \frac{1}{2} d(v^2_{ij})\, = \,- \frac{G \, m_j}{r^2_{ij}}\, dr_{ij}  + \kappa(t) \, v^2_{ij} dt \, ,
\end{equation}
\noindent
Summing on all the interactions of the $j \neq i$ stars with the one noted $i$, then summing on all the $i$ stars, we get
\begin{equation}
  \frac{1}{2} \sum _i \sum_{j \neq i} d(v^2_{ij})\, = \,- \sum_i  \sum_{j \neq i}  \frac{G \, m_j}{r^2_{ij}}\, dr_{ij}+
   \sum _i \sum_{j \neq i}  \kappa(t) \, v^2_{ij} dt \, ,
\end{equation}
\noindent
The square of the velocity of the $i$ star due to its interactions with all the $j$ ones is   $  \frac{1}{N} \sum_{j \neq i} v^2_{ij}$.
In the  double summations, each of the $N(N-1)/2$ interactions is counted twice. To get the mean values, we divide  the double summations
 by $N(N-1)$, ($N$ being very large,  $N(N-1) \approx N^2$) and obtain
\begin{equation}
d(\overline{v^2}) \, \approx \,   - q' \frac{GM}{R^2} dR+ 2  \, \kappa(t) \overline{v^2} dt \, ,
\label{imd3}
\end{equation}
\noindent
where  $M$ and $R$ are the  mass and radius of the configuration 
and $q'$ is a structural factor of the order of unity,   as discussed below.
The terms  $ \frac{G \, m_j}{r^2_{ij}}\, dr_{ij}$ is   scale invariant 
since the effects of a change of scale act the same way on masses and radii. The above expression is then
integrated between time $t_{\mathrm{in}}$, the time of the galaxy  formation
and the present time $t_0$. (As the  analyzed galaxies  have small redshifts, it is like if 
 they are observed at  present.)
 
 At some stage  in the process of galaxy formation,  the infall  of matter 
   was locally  stopped by the centrifugal or pressure forces. 
   Let  us call $t_{\mathrm{in}}$ the time when this state of quasi-equilibrium occured. 
   At this time, since the infall velocity vanishes, the effects of the slow secular term (which depends on the velocity) become
   negligible in the balance between the centrifugal force  and  Newtonian attraction. 
   (We may however notice that this stage of quasi-equilibrium with a vanishing infall velocity does likely not occur
   at the same time for  all  galactocentric radii. However, in the present estimate we do not account for this fact, considering that 
   the spread in the formation times at different radii is small with respect to the age of the galaxies. Moreover, as 
   the present derivation  is only applied   to the dwarf spheroidals (Section 3.2), the  approximation appears rather justified.)
  With  account of these remarks, 
after integration we get the following energy equation,
  \begin{equation}
  \overline{v^2} \, \approx  \, q' \frac{G \, M}{R} +   2 \, \int^{t_0}_{t_{\mathrm{in}} } \kappa(t) \,\overline{ v^2(t)} \, dt \, .
  \label{vir}
  \end{equation}
  \noindent
  The value of $ \overline{v^2} $  is the  average  of the present  velocities,
  $M$ and $R$  are the present values.
   We see that the non-Newtonian term  on the 
  right side leads to   cumulative contributions over the ages. This  effect is not present in the models of cold dark matter
  ($\Lambda$CDM model) and this  may thus give some tests of the different theories (Section \ref{discr}).
   
 

    \begin{figure}[t!]
\centering
\includegraphics[angle=0.0,height=6.5cm,width=10.5cm]{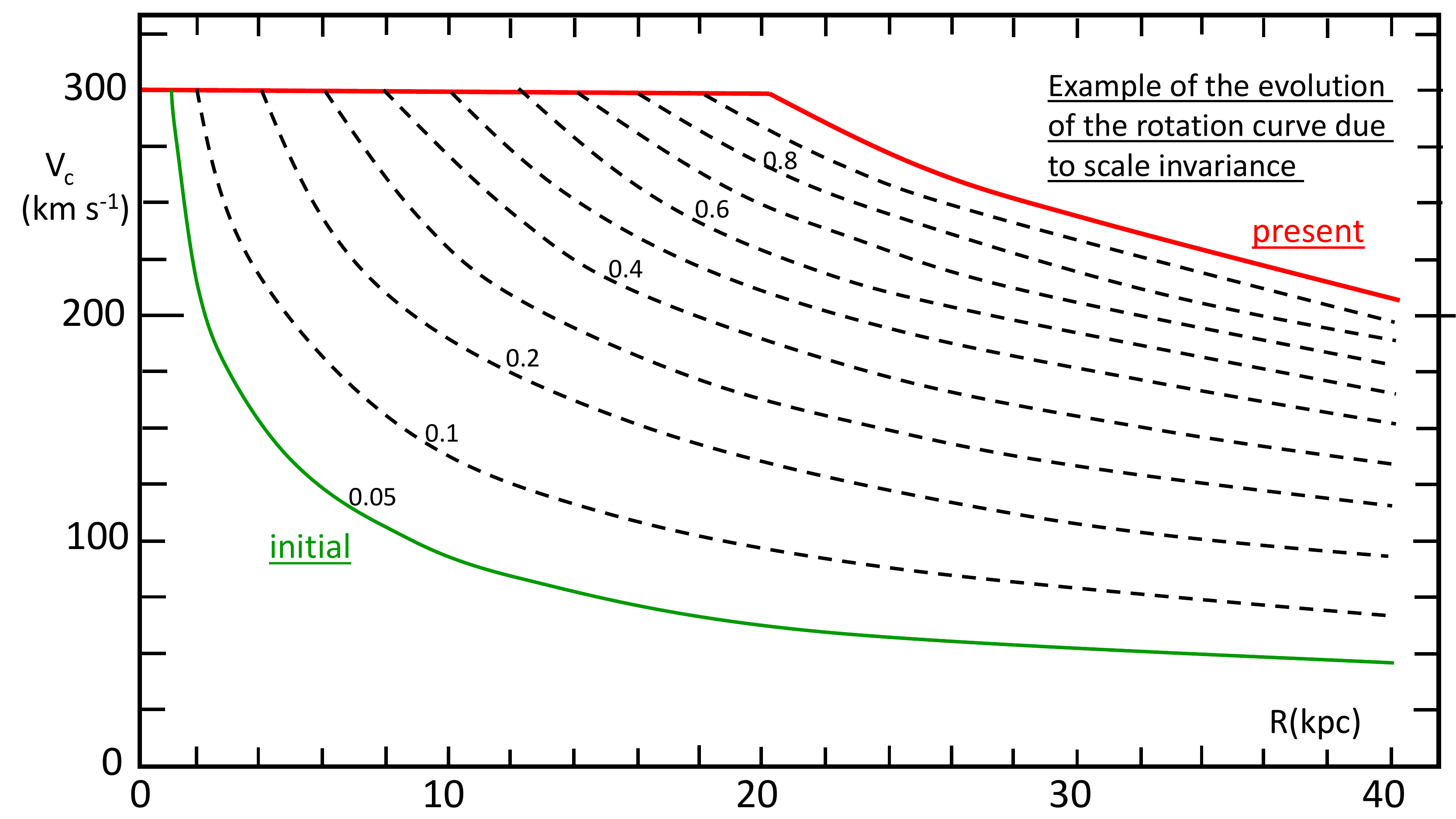}
\caption{Example of the evolution of the rotation curve of a galaxy  in the scale invariant theory.
The initial curve in green corresponds to to a time $t=0.05 \, t_0$, the present curve in red  is that for $t_0=1$.
At a given value of the radius $R$ in kpc,  the square of the circular  velocity is defined by  Equation (\ref {velocity2}). 
        }
\label{ROT}
\end{figure}

 Let us try to express  the above equation  as a  function of the gravities used by  \citet{Lelli17}.
  The approximation is  made,
  \begin{equation}
 2 \,F \, \equiv  \, 2 \, \int^{t_0}_{t_{\mathrm{in}} }\kappa(t) \, \overline{v^2(t)} \, dt \,\approx
  \, 2 \, f  \, \frac{\overline{v^2(t_0)}}{t_0} \, \Delta t \, .
  \label{ff}
  \end{equation}
  \noindent
  The factor $f$ is a numerical factor of the order of unity, discussed below.
  The interval of time  $\Delta t$ is equal to the total distance $\Delta r$ covered by a typical
  star during its circular motions at a constant circular velocity $\overline{v_{\mathrm{circ}}}$,
  thus $\Delta t \approx \Delta r/ \overline{v_{\mathrm{circ}}}$.  The total distance $\Delta r$  is given by the number $n$ of 
  circular orbits that a representative star has described during its lifetime, thus
   $\Delta r  \approx  n \times  2 \, \pi \, \overline{R}$, where  $\overline{R}$  is a time  average of the typical radius. 
  The values of $n$ may be  of the order of  a few tens. 
The present circular velocity is $v_{\mathrm{circ}} \, = \, (g_{\mathrm{obs}} \,R)^{1/2}$.
  According to Equation (\ref{velocity2}), it is   an invariant.
  The  approximation $\overline{R}   \approx (1/2) R$ 
  is taken. This leads to
   \begin{equation}
 2 \,F \, \approx  \, 2  f  \, \frac{\overline{v^2}}{t_0} \, \frac{  2  \pi n  \overline{R}}{(g_{\mathrm{obs}} \, R)^{1/2}} \,
 \approx \,  2  f  \, \frac{\overline{v^2}}{t_0} \, n   \pi  \left(\frac{R}{g_{\mathrm{obs}}} \right)^{\frac{1}{2}} \, .
 \label{ratio}
  \end{equation}
  \noindent 
 The energy equation (\ref{vir}) then  becomes
   \begin{equation}
  g_{\mathrm{obs}} \, \left[1 - \frac{2 f n \pi}{t_0} \left(\frac {R}{g_{\mathrm{obs}}}\right)^{\frac{1}{2}} \right] \, 
  \approx \ q'\, g_{\mathrm{bar}} \, .
  \label{grav}
  \end{equation}
\noindent
  The mean gravitational acceleration $\frac{G \, M}{R^2}$  at some galactocentric distance $R$  encompassing a mass $M$ gives  $g_{\mathrm{bar}}$.
  The estimate of $g_{\mathrm{obs}}$ is based on  $  \overline{v^2} /R $,   the observed  
  velocity being  obtained by the projection of the total velocity.

 This  relation contains  several approximations.
   1.- The first is the structural factor $q'$.  
    The value of $q'$ is   $q' =6/5$ for a constant spherical distribution 
   and $q'=4/3$ for a  disk of constant density. 
   For an almost linearly  regularly decreasing density with a polytropic index $n=1$, 
    $q'=3/2$. 
    Uncertainties on $q'$ may affect the values of $\log g_{\mathrm{bar}}$, however they are at most
      of the order  of one or two tenths of dex.  In fact, this uncertainty disappears in the plot, since
    an appropriate mass-luminosity ratio is applied  so that the observed 
    $\log g_{\mathrm{bar}}$ and $\log g_{\mathrm{obs}}$ are equal for  the highest gravities  in the galaxy sample by \citet{Lelli17}.  
    
 2.-  Another uncertainty concerns the  factor $f$,  expressing  
 the mean of $\frac{v^2}{t}$ over the galaxy lifetime $\Delta t$. The main effect of scale invariance 
 is the progressive radial expansion of a rotating galaxy keeping a constant circular velocity $v_{\mathrm{circ}}$.
 Fig. \ref{ROT} gives an example of a dynamical evolution consistent with the above equations, (see Fig. 2 in \citet{Maeder17c}).
 At a given radius $r$,  the circular velocity at time $t$ is given by
 \begin{equation}
 v_{\mathrm{circ}}(t) \, = \, v_0   \left(  \frac{r_{\mathrm{flat}}(t_{\mathrm{in}})}{r} \, \frac{t}{t_{\mathrm{in}}} \right)^{1/2} \, ,
 \label{vcirc}
 \end{equation} 
 \noindent
 where $r_{\mathrm{flat}}(t_{\mathrm{in}})$ is the radius up to which
  the velocity is about constant ($v_0$) at the initial time  $t_{\mathrm{in}}$.
 The velocity $ v_{\mathrm{circ}}(t) $ cannot be larger than $v_0$.
 At a given  $r$, the value of $ v^2_{\mathrm{circ}}$ grows linearly with time until it reaches $v^2_0$.
 The integral $\int^{t_0}_{t_{\mathrm{in}} } \kappa(t) \, \overline{v^2(t)} \, dt  =v^{2}_0 \left( \frac{t_-t_{\mathrm{in}}}{t_0}\right)$
 and thus  is smaller than 1. If $t_0 > \frac{r}{r_{\mathrm{flat}}} t_{\mathrm{in}}$ the velocity is constant and the integral
  becomes  slightly  larger than 1. On the whole, $f \approx 1$ appears an acceptable approximation.
  
 3.- The mean  radius $\overline{R}$ of an orbit during the galactic lifetime has also been approximated. 
 As the radius of an orbit grows linearly in time, the mean radius over the time $(t_0 - t_{\mathrm{in}})$
 is  $ \overline{R}= 0.55\; R$ or $0.525$, if  the initial time is equal to   10\%  or 5\% of the present time respectively.
 Thus, the   approximations  made appear acceptable in this  analytical approach of a complex problem to be
 further studied by detailed numerical computations. \\

\begin{figure}[t!]
\centering
\includegraphics[angle=0.0,height=6.5cm,width=8.9cm]{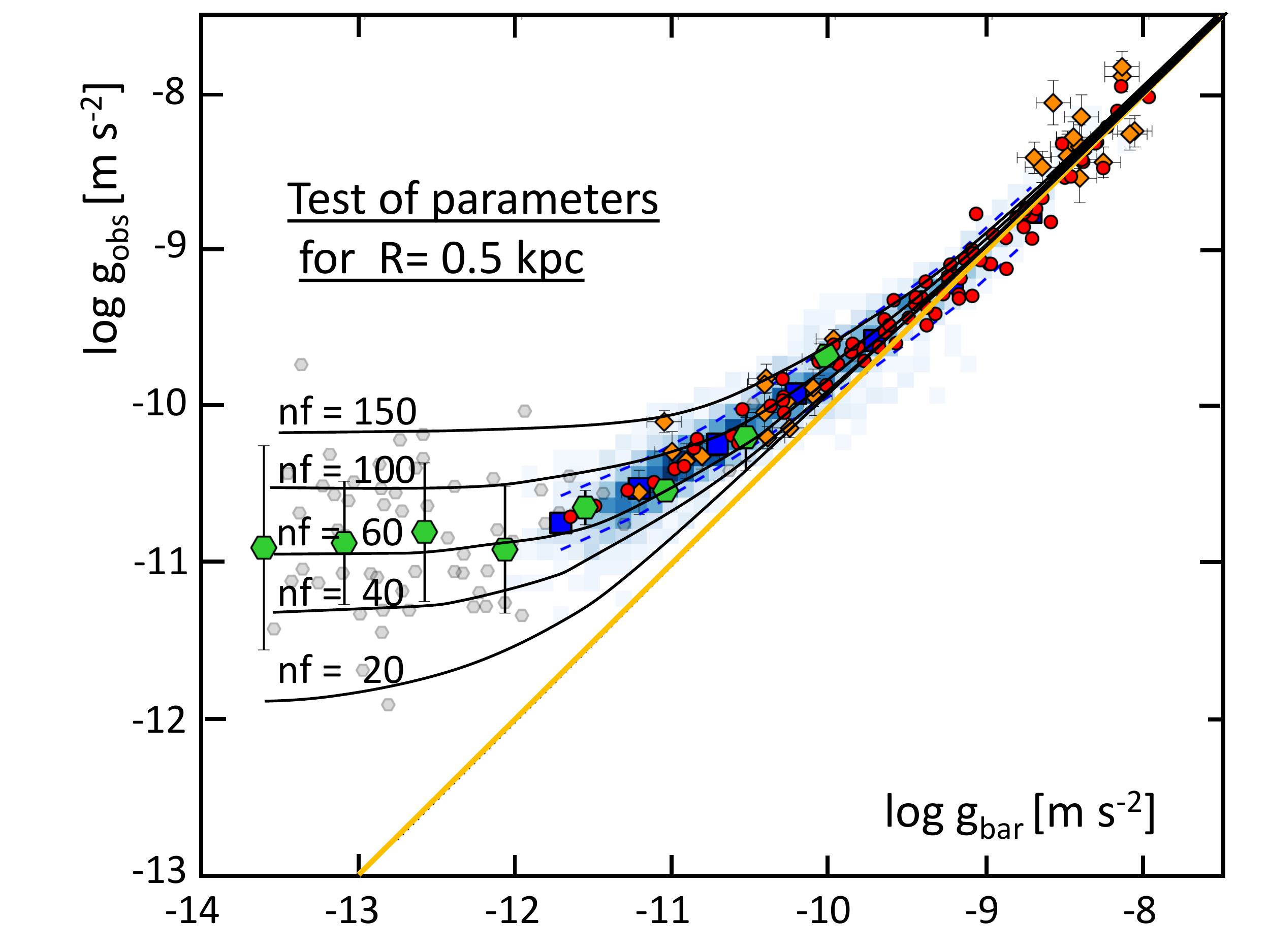}
\includegraphics[angle=0.0,height=6.5cm,width=8.9cm]{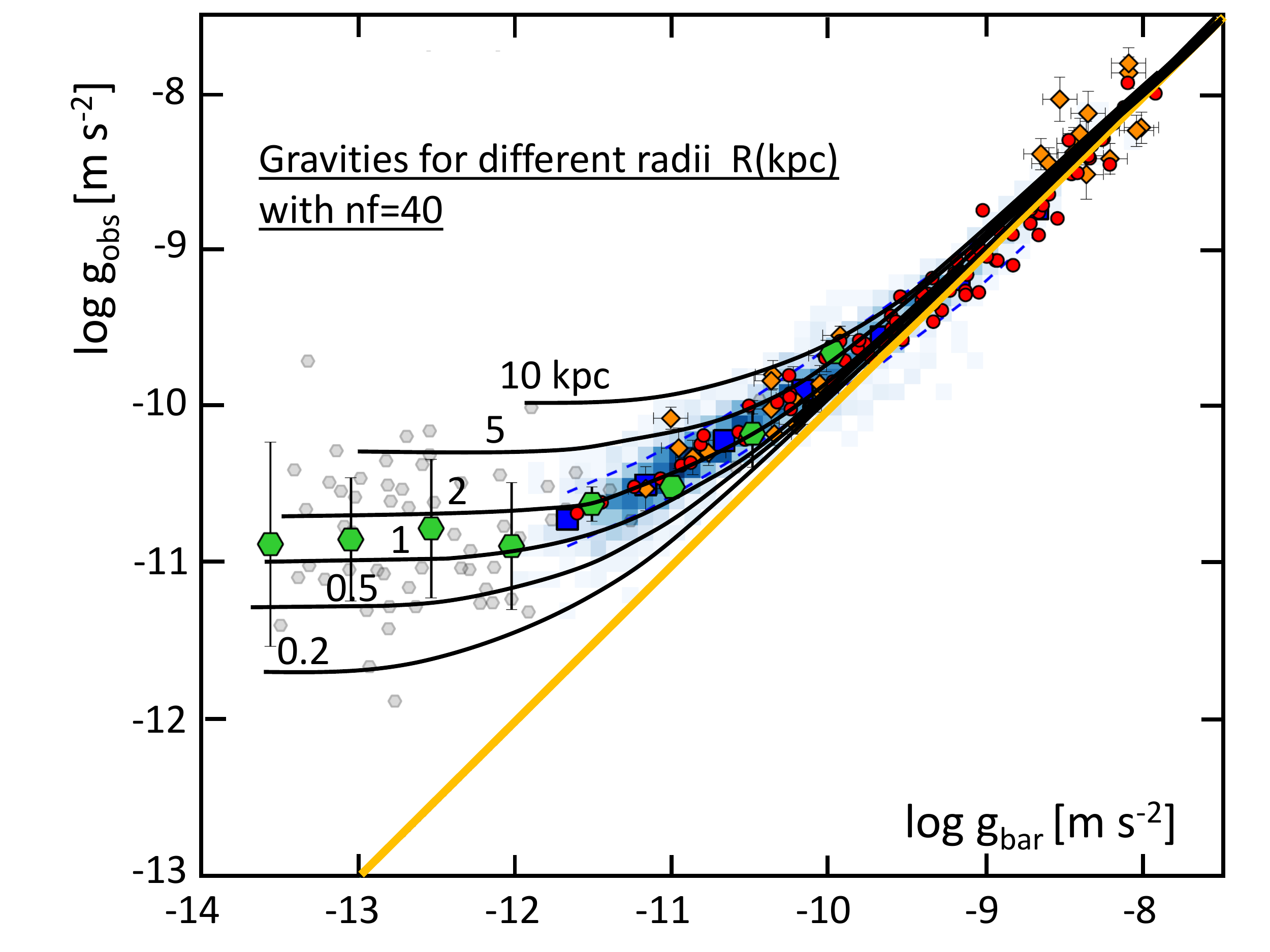}
\caption{Left:The acceleration relation for  all galaxies studied from Fig. 12 left by \citet{Lelli17}, with  continuous black lines 
derived from Equation (\ref{grav}) for different values  of the product $nf$.  
A reference value of $R=0.5$ kpc is taken.
The blue squares 
represent the binned data for 153 LTGs, the thin blue broken lines indicate the standard deviation around their average. The rotating ETGs are represented by red circles, and the X-ray ETGs by orange
diamonds. The small grey hexagons show the dwarf spheroidals, while the large green hexagons give their mean values. Vertical and horizontal bars show standard deviations. The yellow line shows the relation where the two gravities are equal.
Right: the same for $nf=40$ and different radii.
        }
\label{Testnf}
\end{figure} 
 
  Fig. \ref{Testnf} (left) shows the results  for a radius   $R=0.5$ kpc with different values of 
$nf$.
For the high gravities, all the    curves are  close to  the 1:1  yellow line   and thus  the main constraints come
from the low gravity objects which show the largest deviations, in particular the spheroidal galaxies.  
Fig. \ref{Testnf} right  shows the curves   for $nf=60$ and   different radii.
The general distribution  and the flattening for low gravities are well reproduced. 
The flattening results from the fact that for lower  values of $g_{\mathrm{obs}}$
the parenthesis in Equation (\ref{grav}) becomes very small.  
The  predicted  curves tend toward a
 horizontal asymptotic line, in agreement  with 
the general trend discovered by \citet{Lelli17}.  


   In the plot of the RAR of  the LTGs and ETGs,  
 there is a full mixture of  values of $R$ and velocities along the sequence of observations,
 since a  given value of $g_{\mathrm{obs}}$ can be realized by a wide combination of velocities and radii.
 This  prevents  the application of the above developments to this sample.
 However,  in the case of the spheroidals, the situation is more favorable
 since at a given mean $g_{\mathrm{obs}}$ we have a  defined mean radius, as seen in Section (\ref{dSph}).

  \subsection{Relation between gravities and densities} \label{gdens}
 
  For the LTGs and ETGs, we need a relation between $g_{\mathrm{obs}}$ and $g_{\mathrm{bar}}$ which does not depend 
 on the radius or velocity of the observations. 
 The second  term in brackets in Equation (\ref{grav}), which is responsible for the deviations from the 1:1 line in the gravity plot,
 results  from the additional dynamical term in the equation of motion (\ref{Nvec}) integrated over the time.
 The ratio  $x$ at a given time $t$ of the additional term  in Equation (\ref{Nvec}) to the classical Newtonian term behaves like
 \begin{equation}
 x \,= \, \frac{ v \, r^2}{GM \,t} \, \approx \, \sqrt{2} \, \left(\frac{\varrho_{\mathrm{c}}}{\overline{\varrho}} \right)^{1/2} \, ,
 \label{XX}
 \end{equation}
 \noindent
 as shown by \citet{Maeder17c}. 
 There $\overline{\varrho}$ is the mean density of the considered configuration and $\varrho_{\mathrm{c}}$ the classical
 critical density of the Universe at time $t$. The ratio $x$ is a scale invariant term.
 For a disk galaxy of thickness $h$, the average density is 
\begin{equation}
 \overline{\varrho} \,=\, \frac{M}{4 \pi R^2 h} \, = \, \frac{g_{\mathrm{bar}}}{4 \pi G h} \, ,
 \end{equation}
 \noindent
 where $M$ is the mass inside radius $R$ and $g_{\mathrm{bar}}$ the local baryonic gravity. 
 For LTGs of a given thickness as considered by \citet{Lelli17},
 the gravities and mean densities are proportional  to each other. Thus, the  expression of $x$ shows that for LTGs
 the  dynamical effects resulting from the non-Newtonian term
 are proportional   to $g_{\mathrm{bar}}^{-(1/2)}$. 
 
 In  the plot $\log g_{\mathrm{obs}}$ vs. $\log g_{\mathrm{bar}}$, 
 the deviations of  $ \log g_{\mathrm{obs}}$ from the 1:1 line result from these effects integrated over the ages.
 Thus, at two different baryonic
 gravities $g_{\mathrm{bar,1}}$ and $g_{\mathrm{bar,2}}$  for galaxies assumed to be of similar ages, 
 the relative differences  $ \left(\frac{g_{\mathrm{obs}}- g_{\mathrm{bar}}} { g_{\mathrm{bar}}}\right)$
  should  behave like,
 \begin{equation}
   \frac{\left(\frac{g_{\mathrm{obs}}- g_{\mathrm{bar}}} { g_{\mathrm{bar}}}\right)_2}
     { \left(\frac{g_{\mathrm{obs}}- g_{\mathrm{bar}}} { g_{\mathrm{bar}}}\right)_1} \,  =\,  
 \left(  \frac{g_{\mathrm{bar,1}}}{g_{\mathrm{bar,2}}} \right)^{1/2} \, .
   \label{corrg}
  \end{equation}
   \noindent
   Globally the deviations from the 1:1 line in the gravity plot grow like the inverse of  the baryonic gravities with
   a power 1/2. This expression can also be written,
   \begin{equation}
   \left(\frac{g_{\mathrm{obs}}}{g_{\mathrm{bar}}}\right)_2 \, =
    \, 1 + \left[\left(\frac{g_{\mathrm{obs}}}{g_{\mathrm{bar}}}\right)_1 - 1\right]
   \left(\frac{g_{\mathrm{bar,1}}}{g_{\mathrm{bar,2}}}\right)^{1/2}  \, .
  \label{cocorr}
  \end{equation}
  \noindent
  It expresses the changes of the ratio $\frac{g_{\mathrm{obs}}}{g_{\mathrm{bar}}}$ according to the changes of 
  $g_{\mathrm{bar}}$.
  It does not require the values of radius or velocity  and thus it appears of  interest  for the analysis of LTG's galaxies, 
  since they show   a mixture of radii and velocities along the RAR.

 \section{Comparisons between models and observations}
 
 \subsection{Basic calibrations}
 
 The distances of LTGs  in the SPARC sample  \citep{Lelli16} are within the range of a few tens of Mpc. 
 The spheroidals, except  one, are closer 
 than 1 Mpc \citep{Lelli17}. Thus,  one may consider that all these  galaxies are observed at 
 the present time $t_0$. The effective radii
 $R_{\mathrm{eff}}$ encompass the half of the luminosities, for LTGs they range  from about 0.5 kpc to 15 kpc.
  Data for ETGs confirm those for LTGS, while the dSphs extend the curve to the left with a  distinct flattening for the lowest gravities,
 in particular for the ultrafaint spheroidals.
 The  spheroidal galaxies have   radii 
   around 1 kpc for the richest ones,  with values of about 0.2 kpc for the smallest ones.
The complete sample of 240 galaxies  has been represented by \citet{Lelli17} in their Fig. 12 left, this figure is used 
as the basis for observational comparisons.

    \begin{figure}[t!]
\centering
\includegraphics[angle=0.0,height=8.5cm,width=12cm]{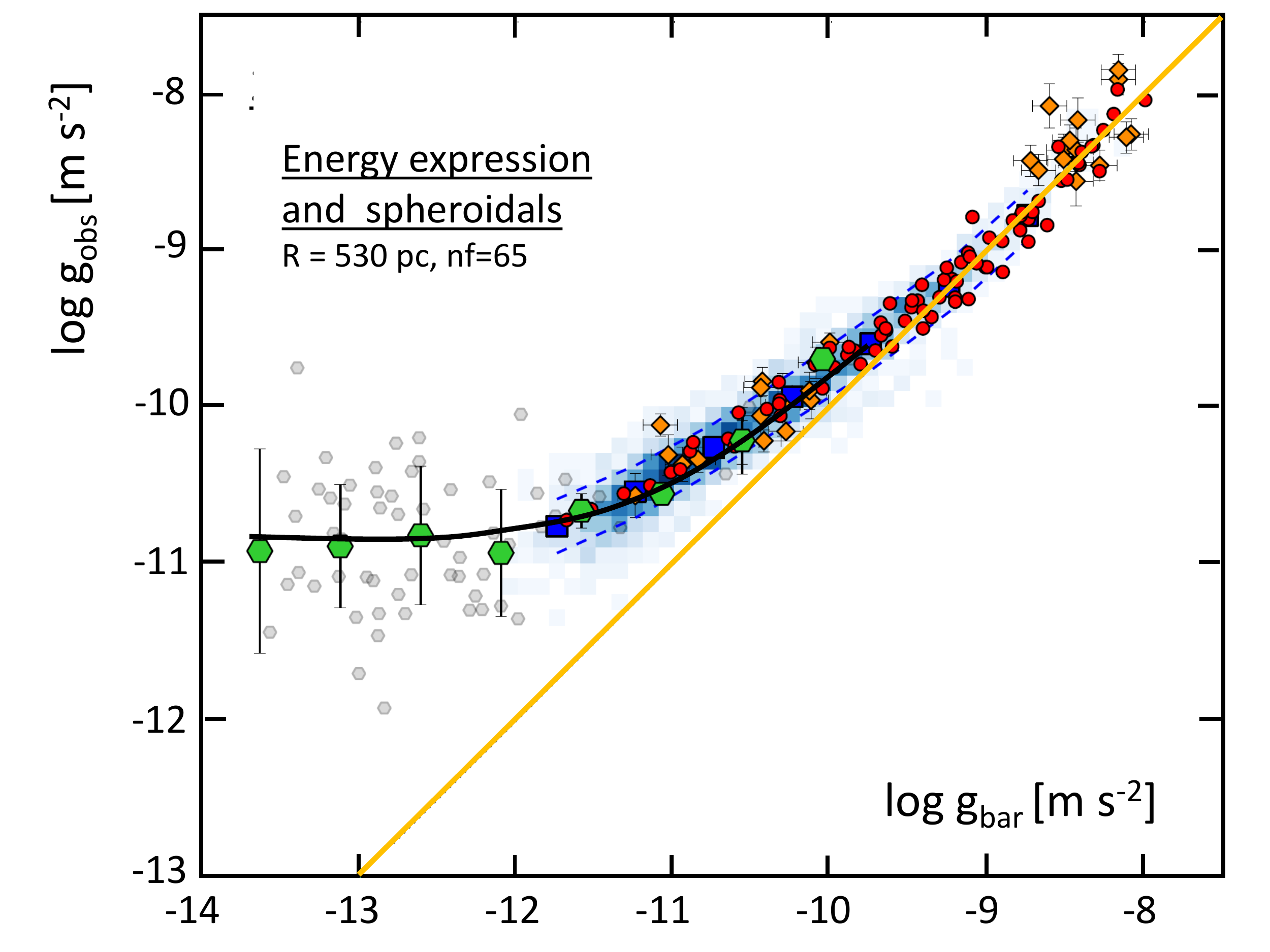}
\caption{The same as Figs. \ref{Testnf}, where the black curve represents the  expression (\ref{grav}) for the spheroidals
for their mean radius of 530 pcs, with $nf=65$. The binned observations for dSphs are the big 
green hexagons.
        }
\label{spheroidals}
\end{figure}

\subsection{The case of the dwarf spheroidal galaxies}  \label{dSph}

The dSphs are the objects where the mass  fraction of dark matter is found to be the highest \citep{Sancisi04}.
The average effective radius of dwarf spheroidals is 530 pcs. Their mean representation in the  $ \log g_{\mathrm{obs}}$
vs.  $ \log g_{\mathrm{bar}}$ plot defines
an horizontal asymptotic line, as suggested by \citet{Lelli17}.  This asymptotic line is located at  $ \log g_{\mathrm{obs}}$
between -10.9 and -10.8. There are only 6  dSphs with  $\log g_{\mathrm{bar}} > -11.50$. Their mean effective radius 
is 539 pcs, about the same as the global average and thus they may be described  by  the same parameters.

Let us now apply expression (\ref{grav})  to the sample of dSphs. We find that  for the mean observed radius $R= 530$ pcs, a value $nf=65$
well  reproduces the observed mean
distribution of the dwarf spheroidal galaxies. This  distribution  corresponds to a predicted  asymptotic line
  at  $ \log g_{\mathrm{obs}}= -10.84$ for the lowest  baryonic gravities,
as illustrated by the black line in Fig. \ref{spheroidals}. 
The above value of  $n$ should correspond to the number of equivalent tours made by a galaxy at the corresponding  
$g_{\mathrm{obs}}$ during its lifetime  $\tau$. This number should be of the order of
$n \, = \, \frac{\tau}{2 \pi} \, \left(\frac{  g_{\mathrm{obs}}}{R} \right)^{1/2}.$
For an age $\tau$ equal to the age of the Universe of 13.8 Gyr  \citep{Frie08},  we get $n=65.2$. The real  age may be 5 or 10\%
smaller, if we account for the formation time.   On the whole, we see that these values  are  consistent.

 Equation (\ref{grav})  describes why  the dark matter, supposed to be responsible for the deviations in the plot,
  is found to be relatively more abundant in low gravity regions and in particular in the 
  dwarf spheroidals. The deviations
 from the 1:1 line negatively depend on the ratio $(\frac {R}{g_{\mathrm{obs}}})^{\frac{1}{2}}$. Over the whole sample,
the mean  $g_{\mathrm{obs}}$ varies by a factor $10^3$, while the mean half light radii  cover a smaller range from 0.5 kpc 
 to  about 15 kpc. Thus,  the variations of gravities largely dominate over the variations of radii in Equation (\ref{grav}).
Thus,  despite their small radii,  the dwarf spheroidals have the largest  term  $(\frac {R}{g_{\mathrm{obs}}})^{\frac{1}{2}}$.

The scale invariant developments well account for the fact that the  dynamical gravities of the dSphs are much higher 
than the baryonic gravities
without calling for some unknown matter component. In addition,  the
convergence of the distribution  of the dSphs and of ultra-faint spheroidals towards an horizontal asymptotic line
in the  plot $\log g_{\mathrm{obs}}$ vs.  $\log g_{\mathrm{bar}}$ is well reproduced. 
The segment of curve corresponding to the slightly higher radii and gravities for dSphs 
with  $\log g_{\mathrm{bar}} > -11.50$ is also  in agreement with  observations.

 \subsection{The late type-galaxies (LTGs)}

    \begin{figure}[t!]
\centering
\includegraphics[angle=0.0,height=8.5cm,width=12cm]{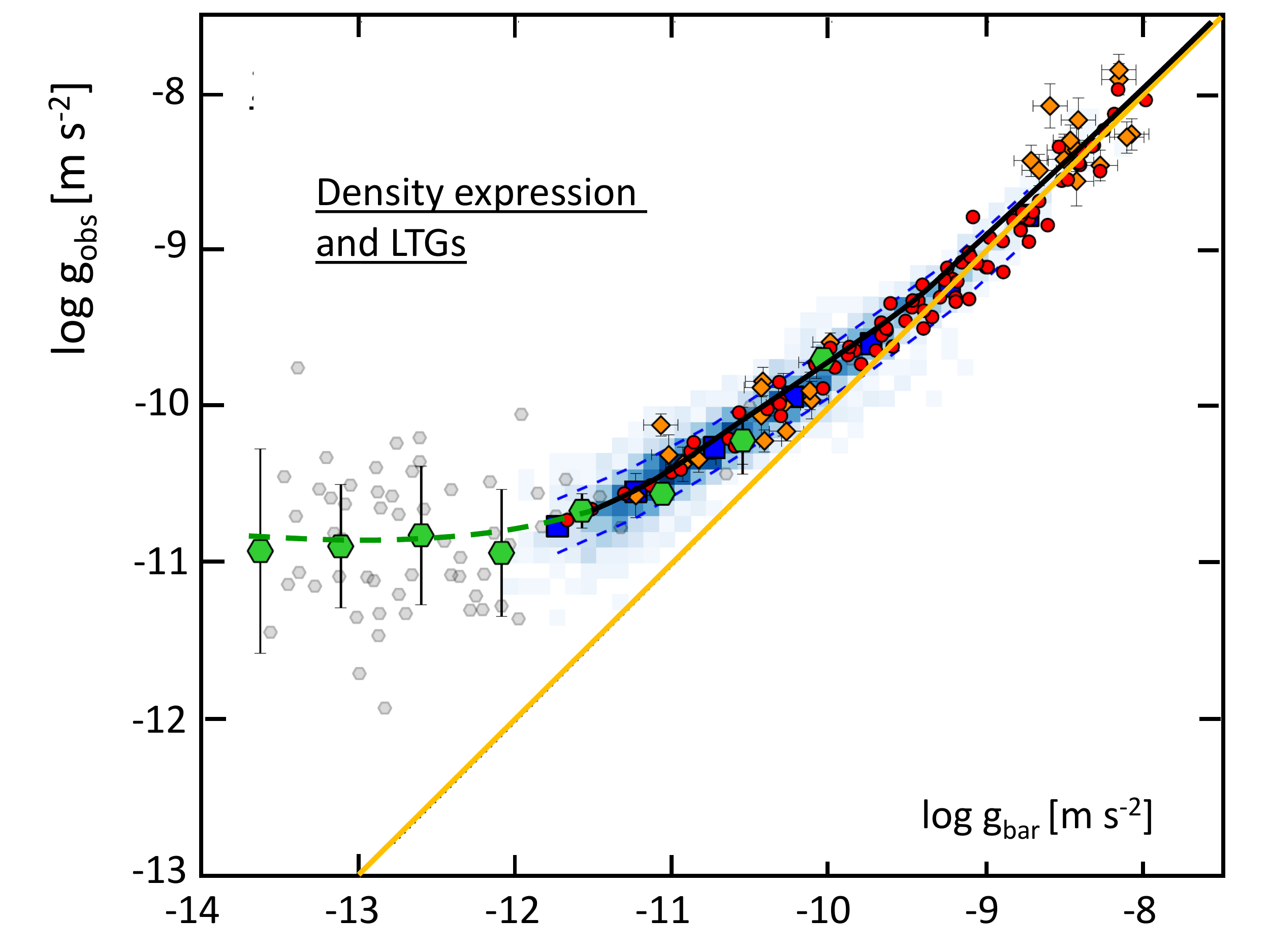}
\caption{The same as Figs. \ref{Testnf} with the results of the scale invariant models. 
The black curve corresponds to  Equation (\ref{cocorr}) applied to LTGs, with an
 adjustment  made by continuity with the spheroidals at $\log g_{\mathrm{bar}} = -11.50$.
The green broken line represents the extension for spheroidals as shown in Fig. \ref{spheroidals}.
        }
\label{overall}
\end{figure}

 As mentioned above, 
along the sequence of LTGs in the  plot $\log g_{\mathrm{obs}}$ vs.  $\log g_{\mathrm{bar}}$, there is a very large variety 
of velocities and radii. As an example, an observation for  a velocity of 240 km/s at a galactocentric distance $R=60$ kpc
has  the same value of $g_{\mathrm{obs}}$ as for  a velocity of 31 km/s at a distance of 1 kpc. 
One cannot associate a given radius to some value of $g_{\mathrm{obs}}$  in the plot for  LTGs,
contrarily  to the case of dSphs. The consequence is that expression (\ref{grav}) is not appropriate  to analyse 
the RAR plot of LTGs.

For these, we rather  apply  the expression of the deviations from the 1:1 line
 given by Equation (\ref{cocorr}).
 As a reference point to perform the adjustment of Equations (\ref{grav}) and (\ref{cocorr}),  we consider  the value $\log g_{\mathrm{bar}}=-11.5$,
  where  the transition from LTGs to spheroidal galaxies  occurs.
  There, Equation (\ref{grav}) gives a value 
  $\log g_{\mathrm{obs}}=-10.695$ (Fig. \ref{spheroidals}).  For a radius of 530 pcs, this corresponds to a velocity
  of 18.7 km/s. This fix the  reference for the application of expression (\ref{cocorr}). (A few spiral galaxies show  observations with about these values of $R$ and
  velocity, in particular NGC 3741, 
  also DD0168, NGC 2455, NCG04278, UGC07524).
  The above  value  of $\log g_{\mathrm{obs}}$ implies  a ratio 
  $\left(\frac{g_{\mathrm{obs}}}{g_{\mathrm{bar}}}\right)= 6.412$ at $\log g_{\mathrm{bar}}=-11.50$.
   Now,  for other  values of $\log g_{\mathrm{bar}}$ (say point 2 in Equation (\ref{cocorr})), 
  the ratios $g_{\mathrm{obs}}/ g_{\mathrm{bar}}$ can be determined. 
 For  $\log g_{\mathrm{bar}}=-11, -10, -9, -8, -7$,   one obtains $\log g_{\mathrm{obs}}= -10.29, -9.64, -8.85, -7.95, -6.98$
 respectively.
  
 Fig. \ref{overall} compares these predicted values  with the observations of  the radial acceleration
  relation for LTGs and ETGs. The agreement appears   satisfactory, with a continuous evolution of the  slope 
 obtained for spheroidal galaxies.
 On the whole,
  we conclude  that  Equations (\ref{grav})  and (\ref{cocorr}), in their respective domains of application,
    reproduce the main features  of 
the  acceleration relation.

\section{Discussion}

Several observed properties of galaxies are direct consequences of the acceleration relation and are thus also consistent with the 
predictions of the scale invariant theory. However, this is also the case for other theories, as discussed in the introduction.
Here, we briefly discuss these points and examine some way to distinguish between the various theories.

\subsection{The Tully-Fisher, Faber-Jackson relations and other properties }

The original Tully-Fisher relation \citep{Tully77} relates the luminosity of spiral galaxies  to the  velocity in 
the flat part of the rotation curve. Further works by \citet{McGaugh00} demonstrate that there is also a well defined  relation
 between the rotation velocity $V$ in the flat part of the rotation curve and the baryonic mass
$M_{\mathrm{bar}}$ of the form  $M_{\mathrm{bar} }  \sim V^4$, this  relation is the so-called BTFR.
  As pointed out by \citet{Lelli17}, the BTFR directly results
from the fact that over a part the observed  acceleration relation does not follow the 1:1 line, but rather behaves
more or less  like
$g_{\mathrm{obs}} \sim g^{1/2}_{\mathrm{bar}}$.  This implies that $V^2 \sim M^{1/2}_{\mathrm{bar}}$, which 
gives a behavior of the baryonic mass  with   about $V^4$.
From
Fig. \ref{overall}, we see that the observed points of LTGs, as well as the scale-invariant 
relation, follow a relation
$g_{\mathrm{obs}} \sim g^{\alpha}_{\mathrm{bar}}$
with $\alpha \approx 1/2$ for  LTGs  below  $\log g_{\mathrm{bar}} < -10$.  For this limited range, the above  scaling of
the velocity with mass is   reproduced. For higher gravities, $\alpha$ increases 
to finally reach  a value of 1 at about  $\log g_{\mathrm{bar}} = -8$, 
 leading to  different slopes of the BTFR relation. These different regimes of the BTFR  specifically
  depend on the mass--radius relation of galaxies
 and their detailed study is beyond the scope of the present work.

The same kind of remarks applies to 
the Faber-Jackson relation \citep{Faber76}, which  relates the luminosity (or the stellar mass) of the elliptical galaxies to their 
velocity dispersion  $\sigma$.  As  shown by \citet{Lelli17},  the facts that 1) the velocity dispersion and the rotation velocities 
of rotating ellipticals are linearly   related,  and 2) that the rotating ETGs follow the same  $M_{\mathrm{bar} }  \sim V^4$
relation as the LTGs  \citep{den Heijer15}  lead to a relation  of the form $M_{\mathrm{bar} }  \sim \sigma^4$.
Also, there is  a close  correspondence  between features in the luminosity profiles and in the rotation curves of galaxies.
As  quoted by \citet{Lelli17}, this  property is known as ''the Renzo rule'' following \citet{Sancisi04}, who emphasizes 
that ''for  any feature in the luminosity  profile there is a corresponding feature in
the rotation curve, and vice versa".
These various  properties  are related to the acceleration relation and consequently to its possible explanation, whatever it is.

\subsection{The various  interpretations of the RAR and the possibility to distinguish them}  \label{discr}

In Figs. 1 to 4, the vertical distance of the observed points above the yellow 1:1 line expresses the ratio 
of  the dynamical (or total) masses  $M_{\mathrm{dyn}}$  to  the  baryonic masses $M_{\mathrm{bar}}$. 
 One has thus the correspondence  $M_{\mathrm{dyn}}/M_{\mathrm{bar}} \approx g_{\mathrm{obs}}/g_{\mathrm{bar}}$ 
 \citep{Lelli17}. Interpreting the deviations  as due to dark matter with mass 
 $M_{\mathrm{DM}}= M_{\mathrm{dyn}}-M_{\mathrm{bar}}$, one has
\begin{equation}
M_{\mathrm{DM}} \, \approx \, \left(\frac{g_{\mathrm{obs}}} {g_{\mathrm{bar}}} -1 \right) \, M_{\mathrm{bar}} \, .
\label{DMest}
\end{equation}
\noindent
 For LTGs and ETGs, 
the ratio of dark to baryonic matter may reach  a factor  of 10, while for dwarf spheroidals
 it ranges from  ten to about 400, with  extreme values up to  1000. 
  A large variety of  particles, known and unknown,
 has been proposed since  more than 30 years to account for the dark matter, see recent reviews by  \citet{Bertone17,Swart17}.
  Several authors have proposed 
   explanations in the framework of the $\Lambda$CDM model of galaxy formation, e.g.   \citet{Navarro17}.
 \citet{Li18} have also found an  agreement of the RAR  with  the MOND theory \citep{Milgrom83,Milgrom09}.
 Indeed,  this is not surprising since the MOND  theory has been initially 
  tailored for explaining the  flat rotation curve of galaxies.
 \citet{Li18} also point out that  the properties of the RAR  remain an open issue for $\Lambda$CDM models of galaxy formation.
As seen above, the scale invariant theory  gives one more interpretation of the RAR, including for the dSphs, often
ignored in other interpretations.

Now, we have to see whether some predictions of the theories  could permit the identification of
   the right one, since among the three different kinds, at least two must not apply.
A signature of  the scale invariant theory resides in the cumulative effects in time. 
This  is well illustrated in the integration over the time of the modified Newton's equation (\ref{Nvec}) performed in Section \ref{enrel}.
In this respect, we note that this cumulative effect is able to account for  the increase  with age of the the ''vertical'' velocity
 dispersion of stars in the Milky Way. If it is further confirmed that high redshift galaxies do not have flat
 rotation curves \citep{Wuyts16,Genzel17, Lang17}, this may indicate that both the vertical and horizontal supports 
 are growing with time. At present, there is no proof of that and  further studies are needed.

\section{Conclusions}

The  study of the   gauging condition in Sect. \ref{gauge}  confirms the relation found earlier between
 the cosmological constant $\Lambda_{\mathrm{E}}$
and the scale factor $\lambda(t)$. An energy relation is obtained which well described the RAR  of the dwarf spheroidal galaxies,
for  mean values of their  radius and $\log g_{\mathrm{obs}}$.  The asymptotic limit found by \citet{Lelli17} is also confirmed.
For spiral galaxies (LTGs),  the  relation between between the local gravity and internal density leads to
an expression of the deviation of the 1:1 line in the  $\log g_{\mathrm{obs}}$ vs.  $\log g_{\mathrm{bar}}$ plot
which agrees with the observations. 

The scale invariant theory allows one to understand  why  the supposed dark matter
is found  concentrated,  with large excesses, in   regions of lower gravities.
In this context, there is no need of dark matter  to account for the  observed acceleration relation.
This is in agreement with previous results \citep{Maeder17c} on  the missing  masses of clusters of
galaxies, on the rotation velocities  of galaxies and on the ''vertical'' velocity dispersion of  stars in the Milky way. 
These positive results encourage further exploration.\\

I express my best thanks to D. Gachet for his  support and to V. Gueorguiev for his support and useful comments.




\end{document}